\documentclass[aps,prd,12pt,amssymb,preprintnumbers,amsmath,amsfonts,nofootinbib]{revtex4-1}
\usepackage{latexsym}
\usepackage{graphicx}

\usepackage{amsbsy}
\usepackage{amstext}
\usepackage{amssymb}
\usepackage{amsmath}

\newcommand{\be}{\begin{equation}}
\newcommand{\ee}{\end{equation}}
\newcommand{\ba}{\begin{eqnarray}}
\newcommand{\ea}{\end{eqnarray}}
\newcommand{\bi}{\begin{itemize}}
\newcommand{\ei}{\end{itemize}}

\usepackage{bm} 

\newcommand{\RR}{{\rm I\kern -.2em  R}}

\newcommand{\pa}{\partial}

\begin{document}

\title{Helicity conservation in  perfect electromagnetic and chiral  fluids}

\author{Cristina Manuel$^1$ and Juan M. Torres-Rincon$^2$}

\affiliation{$^1$Instituto de Ciencias del Espacio (ICE, CSIC),
c.Can Magrans s.n., 08193 Cerdanyola del Vall\`es, Catalonia, Spain, and Institut d’Estudis Espacials de Catalunya (IEEC), c.Gran Capit\`a 2-4, Ed. Nexus, 08034 Barcelona, Spain}

\affiliation{$^2$Departament de F\'isica Qu\`antica i Astrof\'isica and Institut de Ci\`encies del Cosmos (ICCUB), Facultat de F\'isica,  Universitat de Barcelona, c.Mart\'i i Franqu\`es 1, 08028 Barcelona, Spain}

\date{\today}

\begin{abstract}
We derive the total helicity conservation law for a perfect electromagnetic relativistic fluid. As the conservation equation contains the derivative of the magnetic helicity, it can be reshaped as having the same form as the chiral anomaly equation if the fluid is isentropic. We also take the nonrelativistic limit of the helicity conservation law, and check the agreement with the Abanov-Wiegmann equation at zero temperature, but we provide further corrections in the more general case. We then consider chiral fluids, when the chiral anomaly equation has to be incorporated in the hydrodynamical equations, together with other chiral transport effects which exist in the presence of a chiral imbalance. We finally study how  the chiral imbalance modifies the helicity conservation law.

\end{abstract}

\maketitle

\section{Introduction}

Relativistic hydrodynamics has been applied for a long time in a variety of cosmological, astrophysical and nuclear physics scenarios~\cite{rezzolla,Gourgoulhon:2006bn}. The recent experimental program associated to heavy-ion collisions has allowed this effective theory to be further studied and developed~\cite{Romatschke:2017ejr}. More particularly, it has been realized~\cite{Son:2009tf} that quantum effects, such as the so-called quantum chiral anomalies, have to be incorporated in the fluid description of chiral systems, that is, systems made up by massless fermions. As the hydrodynamics contains the expressions of the conservations laws of a system, it seems natural to incorporate the quantum chiral anomaly.  The last might be interpreted as a (quantum) modification of the classical conservation law of the chiral current. A lot of work has been carried out in this modified chiral hydrodynamics, with the description of several new  transport phenomena, that even if originating as quantum effects, have relevant macroscopic effects (see Refs.~\cite{Kharzeev:2013ffa,Kharzeev:2015znc,Landsteiner:2016led} for  reviews and references).

In this article we first study the total helicity conservation law in a relativistic electromagnetic fluid, regardless of its microscopic composition.
We assume that there is no dissipation taking place. We then see that in the isentropic limit, the helicity conservation law takes a rather simple form, and can be written in the same form as the chiral anomaly equation, even if it is a classical effect. It expresses the fact that a combination of the magnetic helicity, the fluid helicity and mixed fluid-magnetic helicity is conserved. This law has been explored and studied in Refs.~\cite{Abanov:2021hio,Wiegmann:2022syo} for barotropic nonrelativistic fluids---see a previous work in Ref.~\cite{Yoshida2014}---later generalized for relativistic systems in Ref.~\cite{Abanov:2022zwm}. Our derivation is more general, and it describes the genuine conservation law for generic isentropic fluids, while we also provide the dynamical helicity evolution valid for baroclinic (non barotropic) fluids. From our relativistic considerations we can take the nonrelativistic limit, showing that we only recover the results of Refs.~\cite{Abanov:2021hio,Wiegmann:2022syo} at zero temperature.

We also consider an electromagnetic fluid made up of massless fermions, that is, a chiral fluid.  Quantum modifications to the hydrodynamics are then considered by including the quantum chiral anomaly. This chiral hydrodynamics can be derived from kinetic theory including quantum corrections in the formalism~\cite{Son:2012zy,Stephanov:2012ki,Chen:2012ca,Manuel:2014dza} (for a recent review see Ref.~\cite{Hidaka:2022dmn} and references therein).
We then study how the helicity conservation law is modified in the presence of a chiral imbalance, when (dissipationless) chiral transport effects have also to be incorporated in the hydrodynamical description of the system. We then see that even in the isentropic limit, the total helicity is not conserved.

The paper is structured as follows. In Sec.~\ref{sec:helcons} we derive the conservation law of total helicity for a charged perfect fluid. In Sec.~\ref{sec:nrlimit} we perform the nonrelativistic expansion of this conservation law, and compare the result with the results of Refs.~\cite{Abanov:2021hio,Wiegmann:2022syo}. In Sec.~\ref{sec:conservchiral} we obtain the total helicity conservation law for a chiral relativistic fluid, thus generalizing the result of Sec.~\ref{sec:helcons} for chiral-imbalanced fluids. In Sec.~\ref{sec:discussion} we discuss possible applications of our equations in the context of cosmology and relativistic heavy-ion collisions. Finally in App.~\ref{app:NR} we detail the non relativistic limit of the Euler equation that leads to the Crocco equation.

Our conventions are such that the metric reads $g^{\mu \nu} = (+,-,-,-)$, and we use natural units $\hbar = k_B = c= 1$ unless otherwise stated. We also re-scale the electromagnetic fields with the electromagnetic coupling constant $e$.

\section{Helicity conservation in relativistic  hydrodynamics~\label{sec:helcons}}

In this section we derive the helicity conservation law for a perfect relativistic charged fluid. Our starting points are the relativistic hydrodynamical equations~\cite{Landau,rezzolla,Gourgoulhon:2006bn}, which are the expressions of the macroscopic conservation laws of the charge current, $j^\mu(x)$, and the  energy-momentum tensor $T^{\mu \nu}(x)$, 

\be \label{consev-laws}
\partial_\mu j^\mu(x) = 0 \ , \qquad
\partial_\mu T^{\mu \nu}(x) + j_\rho(x) F^{\rho \nu}(x) = 0  \ ,
\ee
where $F^{\mu \nu}$ is the electromagnetic tensor.

For a perfect fluid in local equilibrium the charge current and energy-momentum tensor are expressed as
\ba \label{id-bar-flow}
j^{\mu}(x) &=& n(x) \: u^{\mu}(x) \; ,
\\ \label{id-en-mom}
T^{\mu \nu}(x) &=& \bigr[ \epsilon (x)+ P(x) \bigl] \,
u^{\mu}(x) \: u^{\nu}(x)
-P(x) \; g^{\mu \nu} \;,
\ea
where $n(x)$, $\epsilon(x)$ are the particle and energy densities, respectively, while $P(x)$ denotes the pressure. The fluid four velocity $u^\mu(x)$ is normalized as $u_\mu(x) u^\mu(x) =1$. In the following we suppress the spacetime argument of the hydrodynamic variables.
 
In a perfect fluid the entropy current is conserved~\cite{rezzolla}, $\partial_\mu (s u^\mu )= 0$, where $s$ is the (local) entropy density in the fluid rest frame.
This conservation law can be derived by projecting the energy-momentum conservation law along $u_\mu$, and after using the thermodynamic relations,
\be
d\epsilon  =  T ds + \mu d n  \ , \qquad
\epsilon + P  =  Ts +  \mu n  \;.
\ee

The relativistic Euler equation can be obtained by projecting the energy-momentum conservation law onto the direction perpendicular to $u^{\mu}$, resulting in
\be \label{euler1}
(\epsilon + P)\, u^{\nu} \partial_{\nu} u^{\mu} 
= \Delta^{\mu \nu} \pa_\nu P + j_\nu F^{ \mu \nu }   \ , 
\ee
where the projector orthogonal to the fluid 4-velocity is defined as
\be \Delta^{\mu \nu}=g^{\mu \nu} - u^\mu u^\nu \ . \ee

The Euler equation can be converted into the Carter-Lichnerowicz (CL) equation~\cite{Lichnerowicz,Carter,Gourgoulhon:2006bn}. The CL equation reads,
\be
\label{CL1}
n u^\nu [ \partial_\nu ( \mu u_\mu) -  \partial_\mu ( \mu u_\nu) ] +  
s u^\nu [ \partial_\nu ( T u_\mu) -  \partial_\mu ( T u_\nu) ] = n u^\nu F_{\mu \nu} \ .
\ee

For our purposes, it turns out more convenient to express the CL equation in terms of the entropy per particle ${\bar s} = \frac{s}{n} $. Then, the enthalpy per particle reads
\be
h \equiv \frac{ \epsilon +P}{n} = \mu + T {\bar s} \ , \label{eq:enthalpy}
\ee
and we define the enthalpy vorticity tensor as
\be
\Omega_{\mu \nu} \equiv  \partial_\mu ( h u_\nu) -  \partial_\nu ( h u_\mu)  \ .
\ee

The enthalpy vorticity tensor can be related to the vorticity of the system. If we define the dual of any two-rank tensor $W_{\mu \nu}$ as
${\widetilde W}^{\mu \nu} \equiv \frac 12 {\epsilon}^{\mu \nu \alpha \beta} W_{\alpha \beta} $, 
then it is easy to show that the vorticity vector, defined as
\be
\omega^\mu =  \frac 12 {\epsilon}^{\mu \nu \alpha \beta}  u_\nu \partial_\alpha u_\beta \ , \label{eq:omega}
\ee 
is naturally expressed in terms of the dual of the enthalpy vorticity tensor as
\be
\omega^\mu = \frac {1}{2 h} \widetilde {\Omega}^{\mu \nu} u_\nu \ . 
\ee 

Dividing Eq.~(\ref{CL1}) by $n$ we reach to a different form of the CL equation, 
\be
\label{CL-final}
 u^\nu  {\cal F}_{\nu \mu}  = -T \partial_\mu {\bar s}  \ , 
\ee
expressed in terms of a generalized field strength tensor
\be
{\cal F}_{\mu \nu } \equiv  \Omega_{\mu \nu} + F_{ \mu \nu}  \ .
\ee
This generalized strength tensor can be derived from the generalized vector connection,
\be
{\cal A}_\mu \equiv h u_\mu + A_\mu  \ , \label{eq:connection}
 \ee
 as ${\cal F}_{\mu \nu }  = \partial_\mu {\cal A}_\nu - \partial_\nu {\cal A}_\mu $ (and the usual $ F_{\mu \nu }  = \partial_\mu A_\nu - \partial_\nu A_\mu $). In the absence of electromagnetic fields, Eq.~(\ref{CL-final}) simplifies, and gets to a well-known form (cf. Eq.(3.118) of Ref.~\cite{rezzolla}),
 \be
 u^\nu \Omega_{\mu \nu} = T \frac{\pa {\bar s}}{\pa x^\mu} \ ,
 \ee
 that expresses that in the absence of vorticity a relativistic perfect fluid is isentropic $\pa_\mu \bar{s}=0$ (not to be confused with the adiabaticity condition: $u_\mu \pa^\mu \bar{s}=0$~\cite{rezzolla}).

We focus now on the helicity conservation law that is fulfilled in the system. It is convenient to start with   
the Chern-Simons current associated to the electromagnetic fields, which is defined as
\be
J^\mu_{CS} = \frac 12 \epsilon^{\mu \nu \alpha \beta} A_\nu \partial_\alpha A_\beta \ . 
\ee
The zero component of this current  is also known as the magnetic helicity~\cite{Biskamp,Moffattbook}.  Integrated over a closed volume, it is a gauge independent quantity if the  magnetic field lines are either zero or tangential to the associated boundary surface, which is the case we will assume here. It is easy to check that
\be
\partial_\mu J^\mu_{CS} = \frac 14 F^{\alpha \beta} \widetilde{F}_{\alpha \beta}  \ .
\ee

The magnetic helicity measures the linkage and twists of the magnetic field lines in the system~\cite{Biskamp,Moffattbook,Moffatt,Hirono:2015rla}.

The CL equation~(\ref{CL-final}) suggests to define a generalized Chern-Simons current,
\be
{\cal J}^\mu_{CS} \equiv \frac 12 \epsilon^{\mu \nu \alpha \beta} {\cal A}_\nu \partial_\alpha {\cal A}_\beta \ ,
\ee
in terms of the vector connection in Eq.~(\ref{eq:connection}).

This new Chern-Simons current can be written as
\be
{\cal J}^\mu_{CS} =  J^\mu_{CS} + h^2 \omega^\mu +  h B^\mu \ ,
\ee
where we have introduced 
\be B^\mu = \frac 12 \epsilon^{\mu \nu \alpha \beta} u_\nu F_{\alpha \beta} \ . \label{eq:Bfield} 
\ee
Here we have also implicitly assumed that integrated over a closed volume $  \frac 12 \epsilon^{\mu \nu \alpha \beta} h u_\nu F_{\alpha \beta}= 
\frac 12 \epsilon^{\mu \nu \alpha \beta} A_\nu \Omega_{\alpha \beta}$, as after integrating by parts  surface terms can be discarded, assuming
that both the magnetic field and vorticity lines are either zero or tangential to the boundary surface~\cite{Moffattbook,Bekenstein}.

Note that ${\cal J}^0_{CS}$ can be considered as the total helicity of the system, as it  is a combination of the magnetic helicity, the fluid helicity and the mixed magnetic-fluid helicity. Similarly to the magnetic helicity, the fluid and mixed helicities measure the linkage among fluid lines, and fluid and magnetic field lines, respectively.

It is easy to check that the divergence of the new Chern-Simons current satisfies
\be
\partial_\mu {\cal J}^\mu_{CS} = \frac 14 {\cal F}^{\alpha \beta} \widetilde {\cal F}_{\alpha \beta}  \ .
 \ee
Then, after using the identity 
$${\cal F}^{\alpha \beta} \widetilde {\cal F}_{\alpha \beta}  =  
 4 {\cal F}^{\alpha \mu} u_\mu \widetilde {\cal F}_{\alpha \nu} u^\nu  \ , $$
and the CL equation~(\ref{CL-final}), one finds
 \be
\partial_\mu {\cal J}^\mu_{CS} =   T \partial_\alpha {\bar s}  \ \widetilde {\cal F}^{\alpha \beta} u_\beta =  T \partial_\alpha {\bar s}  ( 2 h \omega^\alpha + B^\alpha) \ .  \label{eq:divJCS}
 \ee

If the fluid is isentropic ($\pa_\mu \bar{s}=0$~\cite{rezzolla}), then the right-hand side of Eq.~(\ref{eq:divJCS}) vanishes, and the total helicity conservation law $\partial_\mu {\cal J}^\mu_{CS} =0$ can be expressed as  
\be
  \partial_\mu  ( h^2 \omega^\mu + h B^\mu)=  - \partial_\mu J^\mu_{CS} = - \frac 14 F^{\alpha \beta} \,   \widetilde {F}_{\alpha \beta} = - E \cdot B  \ ,\label{eq:conserhel}
 \ee
where we have defined $E^\mu = F^{\mu \nu}u_\nu$.
 
If the fluid is not isentropic, this equation is corrected as
\be
\label{central-eq}
   \partial_\mu ( h^2 \omega^\mu + h B^\mu)= - E \cdot B  +T (\partial_\alpha {\bar s} ) ( 2 h \omega^\alpha + B^\alpha) \ .
 \ee

This conservation law can also be derived using the hydrodynamical equations obeyed by $\omega^\mu$ and $B^\mu$, as is explicitly shown in Sec.~\ref{sec:conservchiral}. In that derivation we also treat the case of a chiral fluid,  where one has to further consider the anomaly chiral equation associated to the chiral current. 

For an isentropic fluid with vanishing $E^\mu$ and $B^\mu$ one finds,
\be h \pa_\mu \omega^\mu = - 2   \omega^\mu \pa_\mu h 
\ . \ee

In the more general isentropic situation in the presence of electromagnetic fields one can  define the axial current 
\be
j^\mu_A =  h^2 \omega^\mu + h B^\mu \ , 
\ee
it is clear that in the isentropic limit it obeys an equation similar
to the chiral anomaly equation
\be
\partial_\mu j^\mu_A =  - E \cdot B \ .
\ee

The nonrelativistic form of this equation was first discussed by Abanov and Wiegmann  in 
Refs.~\cite{Abanov:2021hio,Wiegmann:2022syo}. In
Ref.~\cite{Abanov:2022zwm} the equation was generalized to the relativistic regime (see Eq.~(51) in that reference), but we see that the Abanov-Wiegmann (AW) equations  are only valid at zero temperature, as only then $h = \mu$, cf. Eq.~(\ref{eq:enthalpy}).

Note also while we have considered a flat Minkowski space, it is possible to generalize all our equations in nontrivial metrics, considering the CL equation in a general background metric~\cite{rezzolla,Gourgoulhon:2006bn,Shi:2022jya}.

\section{Helicity conservation in nonrelativistic hydrodynamics~\label{sec:nrlimit}}

In this section we take the nonrelativistic limit to the helicity conservation equation~(\ref{central-eq}), to deduce the corresponding form of the conservation law. We then check that we reproduce the Abanov-Wiegmann equation in the limit of zero temperature \cite{Abanov:2021hio,Wiegmann:2022syo}, while several corrections are needed otherwise.

The AW equation of helicity conservation read~\cite{Abanov:2021hio,Wiegmann:2022syo},
\be
 \partial_t \rho_{\rm AW} + \bm{\nabla} \bm{j}_{\rm AW}=  2 \bm{E} \cdot \bm{B} \ , \label{eq:AW}
\ee
where ${\bm E}, {\bm B}$ are the electric and magnetic fields, respectively. The density 
$\rho_{\rm AW}$ includes a combination of the fluid and mixed helicities
\be
 \rho_{\rm AW}   =    m \bm{v} \cdot \bm{\omega}_{\rm     AW}  + 2 m \bm{v} \cdot \bm{B} \ ,  
\ee
written in terms of the (unusually normalized) vorticity vector  ${\bm \omega}_{\rm AW} = m \bm{\nabla} \times \bm{v}$, while the current is given by
\be
 \bm{j}_{\rm AW} =   \bm{v}  \rho_{\rm AW}   +  (\bm{\omega}_{\rm AW}  + 2  \bm{B}) (\mu_{\rm nr}  - \frac12 m v^2)    - m \bm{v} \times( \bm{E}  +  \bm{v} \times \bm{B})  \ ,
\ee
where $\mu_{\rm nr}$ is the nonrelativistic chemical potential.

Our goal in this section is to arrive to these expressions by performing the nonrelativistic limit of our equations.

In order to take the nonrelativistic limit of Eq.~(\ref{central-eq}),  we momentarily restore the speed of light constant $c$, while keeping $\hbar = k_B =1$. We define $x^\mu = ( ct, \bm{x})$, while $\partial^\mu = (c^{-1} \partial_t,  - {\bm \nabla})$.
 We have
\begin{align}
u^\mu & =   \gamma ( 1, \bm{v}/c) \ , \qquad \gamma = (1 -v^2/c^2)^{-1/2} \ , \label{eq:4velocity} \\
E^\mu  & =  F^{\mu \nu} u_\nu  =  \gamma \left(  \frac{ \bm{v}}{c} \cdot \bm{E}, \bm{E} + \frac{ \bm{v}}{c} \times \bm{B} \right) \ ,
\label{electric}\\
B^\mu &=  {\widetilde F}^{\mu \nu}u_\nu = \gamma \left(  \frac{ \bm{v}}{c} \cdot \bm{B}, \bm{B} - \frac{ \bm{v}}{c} \times \bm{E} \right)  \ ,\\
\omega^\mu &= \gamma^2 \left( \frac{1}{2 c^2} \bm{v} \cdot (\bm{\nabla} \times  \bm{v}), \frac{1}{2 c}  ({ \bm \nabla} \times  {\bm v}) +\frac{1}{2 c^3}{ \bm{v}} \times  \partial_t \bm{v} \right) \ ,  \\
F^{\mu \nu}   \widetilde { F}_{\mu \nu} &= - 4 \, \bm{E} \cdot \bm{B} \ .
\end{align}

The relativistic enthalpy can be written separating the rest energy density, 
\be
h =
m c^2 +  h_{\rm nr} = mc^2 +  \mu_{\rm nr} + T {\bar s}\ ,  \label{eq:nrchem}
\ee 
so clearly we take into account also that the relativistic and nonrelativistic chemical potentials differ by the rest energy $mc^2$.

Then, in the limit $ v \ll c$ we find
\be
 \rho_A \equiv h^2 \omega^0 +  h B^0 = \frac {m c^2}{ 2} \left[ m \bm{v} \cdot (\bm{\nabla }\times  \bm{v}) + 2 \frac{ \bm{v}}{c} \cdot \bm{B} \right] + {\cal O}\left( \frac{v}{c} \right) \ , \label{eq:rhoA}
\ee
The leading term, neglecting other nonrelativistic corrections, is equal to $c^2\rho_{\rm {AW}}/2$.

If we use the Crocco equation (see Eq.~(\ref{eq:Crocco}) in Appendix~\ref{app:NR}),
\be
m \partial_t \bm{v} + \bm{\nabla} ( h_{\rm nr} + \frac 12 mv^2) -   m \bm{v} \times (\bm{\nabla } \times  \bm{v})  = \bm{E} + \frac{ \bm{v}}{c} \times \bm{B} +  T \bm{\nabla} {\bar s} \ , 
\ee
one can write up to ${\cal O} (v/c)$
\be
\partial_i (h^2 \omega^i  )= \frac{m c}{2} \bm{ \nabla} \left[ \bm{v} ( \bm{v} \cdot \bm{ \omega}_{\rm{AW}} ) + \bm{ \omega}_{ \rm{AW}} \left( \frac{ h_{\rm nr}}{m} -  \frac 12 v^2 \right) + \bm{v} \times \left( \bm{E} + \frac{ \bm{v}}{c} \times \bm{B} \right)+    T \bm{v} \times \bm{\nabla} {\bar s}   \right] \ ,
\ee
written in terms of the $\bm{\omega}_{\rm{AW}}$ vorticity; 
while
\be 
\partial_i (h B ^i  ) = m \bm{\nabla} \left[ \frac12 \bm{v} ( \bm{B} \cdot \bm{v}) - \frac12 \bm{v} \times ( \bm{v} \times \bm{B} ) + \frac{ h_{\rm nr}}{m} \bm{B} - c \bm{v} \times \bm{E} \right]  \ ,
\ee
up to ${\cal O} (v/c)$. In the isentropic case we have
that the leading nonrelativistic correction is given by
\be
\bm{j}_A \equiv h^2 \bm{\omega}+h \bm{B} =   \frac { \bm{v}}{c}  \rho_A   + \frac c 2 \left( \bm{\omega}_{\rm{AW}}  + \frac{2}{c}  \bm{B} \right) \left(h_{\rm nr} - \frac12 m v^2 \right)   - \frac{mc^2}{2} \left[ \frac{ \bm{v}}{c} \times \left( \bm{E}  +  \frac{ \bm{v}}{c} \times \bm{B} \right) \right]   \ ,  \label{eq:jAisen}
\ee
which keeps the same form as $c \bm{j}_{ \rm{AW} }/2$, but replacing $ \mu_{\rm nr}$ by $h_{\rm nr}$. 

Finally, from our Eq.~(\ref{eq:conserhel}) we arrive---for isentropic fluids---to
\be
\frac{1}{c^2} \partial_t \rho_A + \frac{1}{c} \bm{\nabla} \bm{j}_A=  \frac{1}{c} \bm{E} \cdot \bm{B} \ .
\ee

If we multiply the whole equation by $2$ we reproduce the AW equation~(\ref{eq:AW}) if we go to the zero temperature limit, as only in this case $h_{\rm nr}= \mu_{\rm nr}$. Note that in Ref.~\cite{Abanov:2021hio} the thermodynamical relation $dp = n d \mu_{\rm nr}$---which is only strictly valid at zero temperature---was used for the derivation of the helicity conservation law, while here we apply the more general Gibbs-Duhem relation $dp= nd\mu_{\rm nr}+ s dT$. Notice also that since an isentropic fluid is also barotropic~\cite{rezzolla,Gourgoulhon:2006bn}, we are able to arrive to the results of Ref.~\cite{Abanov:2021hio} by setting $\pa_\mu \bar{s}=0$.

If the fluid is not isentropic further corrections to the equation are needed, and we arrive to 
\be
\frac{1}{c^2}\partial_t \rho_A + \frac{1}{c} {\bm \nabla} {\bm j}_A= \frac{1}{c} {\bm E} \cdot {\bm B} +   T \left( {\bm \omega}_{AW} + \frac{1}{c} {\bm B}  \right) \cdot {\bm \nabla}   {\bar s} \ , \label{eq:consjA}    
\ee
which is valid up to order $v^2/c^2$. In the conservation law~(\ref{eq:consjA}), for the non isentropic case, $\rho_A$ is still given by Eq.~(\ref{eq:rhoA}), while the current ${\bm j}_A$ shown in (\ref{eq:jAisen}) receives a correction,
\be
\bm{j}_A = \frac { \bm{v}}{c}  \rho_A   + \frac c 2 \left( \bm{\omega}_{\rm{AW}}  + \frac{2}{c}  \bm{B} \right) \left(h_{\rm nr} - \frac12 m v^2 \right)   - \frac{mc^2}{2} \left[ \frac{ \bm{v}}{c} \times \left( \bm{E}  +  \frac{ \bm{v}}{c} \times \bm{B} \right) \right]  + \frac{Tmc}{2} ( \bm{v} \times \bm{\nabla} \bar{s} ) \ , \label{eq:jAnonisen}
\ee

Let us finally stress that Eq.~(\ref{eq:consjA}) is valid for any perfect electromagnetic fluid in its nonrelativistic regime.

\section{Helicity conservation in chiral relativistic plasmas~\label{sec:conservchiral}}  

In this section we consider a chiral relativistic fluid and obtain the total helicity conservation relation using the equations of motion of $\omega^\mu$, Eq.~(\ref{eq:omega}), and $B^\mu$, Eq.~(\ref{eq:Bfield}).

To begin with, we consider a relativistic fluid defined by its $U(1)$ current vector $j^\mu$, its chiral current $j_5^\mu$, and its energy-momentum tensor $T^{\mu \nu}(x)$. In the presence of chiral imbalance, characterized by a chiral chemical potential $\mu_5$, new dissipationless transport phenomena associated to the chiral anomaly has to be incorporated in the hydrodynamical description \cite{Son:2009tf,Kharzeev:2013ffa}. We still work in the perfect fluid (nondissipative) situation, but we need to add corrections to the vector and chiral currents proportional to $B^\mu$ and $\omega^\mu$. In this respect we use the Landau reference frame which allows us to include these corrections only in $j^\mu$ and $j_5^\mu$, but not in the energy-momentum tensor~\cite{Son:2009tf,Sadofyev:2010pr,Landsteiner:2016led}. Therefore, $T^{\mu \nu}$ is still given by the expression in Eq.~(\ref{id-en-mom}), but the currents are now given by~\cite{Sadofyev:2010pr} 
\begin{align}
 j^\mu & = nu^\mu + \xi \omega^\mu +\xi_B B^\mu  \ , \label{eq:jvec} \\
j^\mu_5  & = n_5 u^\mu + \xi_5 \omega^\mu + \xi_{B,5} B^\mu  \ , \label{eq:jaxi}
\end{align}
where $\omega^\mu$ and $B^\mu$ are defined in Eqs.~(\ref{eq:omega}) and (\ref{eq:Bfield}), respectively. The scalars $\xi, \xi_B, \xi_5, \xi_{B,5}$ are the coefficients describing the chiral vortical, chiral magnetic, axial vortical, and chiral separation effects, respectively~\cite{Sadofyev:2010pr,Kharzeev:2015znc}. 

Having both vector and chiral densities, the thermodynamic relations should be generalized. To accommodate both densities into the specific enthalpy density $h$ we now take,
\be h = \frac{\epsilon+P}{n} = \frac{sT+n\mu + n_5 \mu_5}{n} \ , \ee
We also apply the following thermodynamic relation in terms of the specific entropy $\bar{s}$ and the chiral fraction density $x_5=n_5/n$,
\be
dP =ndh-Tnd \bar{s} - \mu_5 n dx_5 \ . \label{eq:dp} 
\ee

The local conservation equations for hydrodynamics read (\ref{consev-laws}), together with the additional chiral anomaly equation,
\be
\pa_\mu j_5^\mu   = -C E_\mu B^\mu \ , \label{eq:consercurrent5}
\ee
with $C=1/(2\pi^2)$ the anomaly coefficient.

From the conservation of the energy-momentum tensor we find two independent equations. Projecting it along $u^\mu$ one finds the usual,
\be D \epsilon + (\epsilon +P) \pa_\mu u^\mu  = 0 \ ,
\ee
where $D \equiv u_\mu \pa^\mu$ and $u_\nu \pa_\mu u^\nu=0$ has been used.

The second equation is obtained by projecting the conservation of the energy-momentum tensor by the projector $\Delta_{\mu \nu}$, arriving to a generalization of Eq.~(\ref{euler1}),

\be (\epsilon + P) u^\nu \pa_\nu u^\mu - \Delta^\mu_{ \ \nu} \pa^\nu P  = \Delta^\mu_{ \ \nu} (n E^\nu  + \xi F^{\nu \lambda} \omega_\lambda + \xi_B F^{\nu \lambda} B_\lambda) \ ,
\label{eq:Euler}
\ee
where we keep the projector in the right-hand side for convenience (even when $\Delta^{\alpha}_{\ \nu} F^{\mu \nu}= F^{\mu \alpha}$). Equation (\ref{eq:Euler}) can be seen as a generalization of the relativistic Euler equation with electromagnetic fields in the presence of chiral imbalance. 

To obtain the equation of helicity conservation we start by considering the divergence of the vorticity,
\be \pa_\mu \omega^\mu = \frac12 \epsilon^{\mu \nu \lambda \rho} \pa_\mu u_\nu \pa_\lambda u_\rho \ , \label{eq:divomega}  \ee
which follows from the definition of $\omega^\mu$ in Eq.~(\ref{eq:omega}). The right-hand side of this equation can be rewritten as,
\be
\frac12 \epsilon^{\mu \nu \lambda \rho} \ \pa_\mu u_\nu \pa_\lambda u_\rho =-\epsilon^{\mu \nu \lambda \rho} \ u_\nu \pa_\lambda u_\rho \ u^\alpha \pa_\alpha u_\mu =-2 \omega^\mu u^\nu \pa_\nu u_\mu \ . \ee

Inserting this relation into Eq.~(\ref{eq:divomega}) we can then use the Euler equation (\ref{eq:Euler}) to obtain,
\be \pa_\mu \omega^\mu = -\frac{2}{\epsilon + P}   \omega^\mu (\pa_\mu P + n E_\mu)   \ , \label{eq:divomega2} \ee
where we have used that $B_\mu F^{\mu \lambda} B_\lambda = \omega_\mu F^{\mu \lambda} \omega_\lambda =\omega_\mu F^{\mu \lambda}B_\lambda=0$.
Equation (\ref{eq:divomega2}) (up to a change of sign in the last term due to metric convention) was also derived in Refs.~\cite{Son:2009tf,Neiman:2010zi}.

On the other hand, from the definition of the  field in~(\ref{eq:Bfield}), we calculate its divergence
\be 
\pa_\mu B^\mu = \frac12 \epsilon^{\mu \nu \alpha \beta} \pa_\mu u_\nu F_{\alpha \beta} + \frac12 \epsilon^{\mu \nu \alpha \beta} u_\nu \pa_\mu F_{\alpha \beta} \ , \label{eq:divB}
\ee
where the last term vanishes identically.

The first scalar term in Eq.~(\ref{eq:divB}) is simplified by the following relation,
\be 
\frac12 \epsilon^{\mu \nu \alpha \beta} \ \pa_\mu u_\nu F_{\alpha \beta} = \frac12 \epsilon^{\mu \nu \alpha \beta} F_{\alpha \beta} u_\mu u^\lambda \pa_\lambda u_\nu +\epsilon^{\mu \nu \alpha \beta} u_\nu \pa_\alpha u_\beta F_\mu^\lambda u_\lambda \ .
\ee

Therefore we end up with
\be
\pa_\mu B^\mu  =  2 \omega \cdot E - \frac{1}{\epsilon + P}  (B \cdot \pa P + n E \cdot B) \ , \label{eq:divB2}
\ee
which coincides with the expression in Refs.~\cite{Son:2009tf,Neiman:2010zi} (up to a sign in the terms carrying $E^\mu$ due to metric convention).

Now we manipulate the Eqs.~(\ref{eq:divomega2}) and (\ref{eq:divB2}), by multiplying them by $h^2$ and $h$, respectively, 
\be
\left\{ 
\begin{array}{ccc}
h^2 \pa_\mu \omega^\mu & = & - 2 h   \omega \cdot \pa h +2 h T  \omega \cdot \pa \bar{s}  +2h \mu_5 \omega \cdot \pa x_5-2h \omega \cdot E   \\
h \pa_\mu B^\mu  & = &  2 h \omega \cdot E - B \cdot \pa h + T B \cdot \pa \bar{s}  +\mu_5 B \cdot \pa x_5-  E \cdot B
\end{array}
\right.
\ee
where we also applied the thermodynamical relation~(\ref{eq:dp}).

Rewriting the two equations as
\be 
\label{hydroeqsmu5}
\left\{
\begin{array}{ccc}
\pa_\mu (h^2  \omega^\mu) & =& 2h T \omega \cdot \pa \bar{s}  +2h \mu_5 \omega \cdot \pa x_5 -2 h \omega \cdot E \ ,   \\
\pa_\mu (hB^\mu)  & = & 2 h \omega \cdot E + T B \cdot \pa \bar{s} +\mu_5 B \cdot \pa x_5 -  E \cdot B \ ,
\end{array}
\right.
\ee
it is easy to see that their sum gives
\be
\pa_\mu (h^2  \omega^\mu+hB^\mu)
= - E\cdot B + (2h \omega^\mu + B^\mu) (T\pa_\mu \bar{s}+\mu_5 \pa_\mu x_5) \ , 
\label{eq:conswithmu5}
\ee
where $E \cdot B=\frac14 \widetilde{F}^{\mu \nu} F_{\mu \nu}$. Notice that in the absence of chiral imbalance $x_5=0$ the expression obtained in Eq.~(\ref{central-eq}) is recovered. 

Equation (\ref{eq:conswithmu5}) is one of the main results of this work. It describes the (non) conservation of the total helicity of a chiral non isentropic fluid. Even when $\partial_\mu \bar{s}=0$ (for example in barotropic fluids), the total helicity is not conserved due to the presence of chiral imbalance. The later itself evolves according to the chiral anomaly equation, which reads 
\be
\pa_\mu (n_5 u^\mu + \xi_5 \omega^\mu + \xi_{B,5} B^\mu) = - C E \cdot B \ . \label{eq:anomaly}
\ee
This equation coincides with the conservation law of Refs.~\cite{Avdoshkin:2014gpa,Yamamoto:2015gzz}. In the last reference this equation is called total ``helicity'' conservation. However, given our definitions, and following the terminology of Ref.~\cite{rezzolla} the (non)-conservation of the helicity is given by~(\ref{eq:conswithmu5}) instead, while we simply  refer to Eq.~(\ref{eq:anomaly}) as the chiral anomaly equation.

The chiral anomaly equation has already been used to see the possible transfer of chirality and helicity among the different sectors of the system \cite{Avdoshkin:2014gpa,Yamamoto:2015gzz,Avkhadiev:2017fxj,Kirilin:2017tdh}. For the consistent description of the magnetic helicity evolution we would need to consider dynamical electromagnetic fields and couple the Maxwell equations~\cite{Boyarsky:2011uy,Manuel:2015zpa}
\be 
\pa_\mu F^{\mu \nu}=e j^\nu \quad , \quad \pa_\mu \widetilde{F}^{\mu \nu}=0 \ .
\ee
This is left for future publications. In the present work we stress that if there is a conserved current, Eq.~(\ref{eq:conswithmu5}) has also to be taken into account.

\section{Discussion and Summary~\label{sec:discussion}}

The helicity conservation law that we have derived for perfect relativistic fluids might lead to several different effects. Depending on the physical situation, the helicity could be transferred from the fluid to the electromagnetic sector and/or viceversa. It remains to be studied how the inclusion of dissipative effects might alter such a possibility, as viscosities and electrical conductivity will affect the dynamical evolution of the helicity. We leave these studies for future projects, as a well-defined
relativistic formulation of the hydrodynamical effects is not
straightforward. Further, the back-reaction of the electromagnetic fields should also be considered.

Note that we have considered a situation quite different from the one of ideal magnetohydrodynamics, when one typically assumes that the electrical conductivity is infinite, which then imposes the constraint $E^\mu =0$ (see Eq.~(\ref{electric})). In that case, the magnetic helicity is conserved, and if we further take $x_5=0$ in Eqs.~(\ref{hydroeqsmu5}), these tell us that in the isentropic limit both the fluid and mixed magnetic helicity are conserved independently~\cite{Bekenstein}, and thus there is no possibility of  transfer of helicity among the different sectors of the system.

We believe that our considerations might be relevant in different cosmological scenarios, as the chiral anomaly plays a central role in different baryogenesis and leptogenesis models.
It might also be relevant for the physics associated to heavy-ion collisions. The fluid created  in these experiments is known to be almost perfect,
and thus, ignoring the role of viscosities might be a good approximation.
Further, it is known that a large vorticity is generated in these experiments~\cite{STAR:2017ckg,Huang:2020dtn}, while different transport codes describing these experiments reveal that  fluid helicity is also recreated \cite{Tsegelnik:2022eoz}. On the other hand, large magnetic fields are also generated in the collision \cite{Kharzeev:2015znc}, while one might expect also the generation of chiral imbalance. This last point has triggered substantial theoretical and experimental efforts for the search of chiral effects in heavy-ion collisions~\cite{Zhao:2019hta,STAR:2021mii}. However, we note that the presence of a chiral imbalance also modifies the entropy content of the fluid (even in the dissipationless case)~\cite{Son:2009tf,Sadofyev:2010pr}, and, according to our Eq.~(\ref{eq:conswithmu5}), the initially-generated chiral density can be transferred not only to magnetic helicity~\cite{Manuel:2015zpa,Tuchin:2017vwb,Rogachevskii:2017uyc,Mace:2019cqo}, but also to fluid helicity. We plan to investigate more on this interplay in future publications by considering dynamical electromagnetic fields reacting to the presence of the fluid’s helicity via the Maxwell equations.

\section*{Acknowledgements} 

We thank A. Abanov for bringing Ref.~\cite{Yoshida2014} to our attention. We have been supported by Ministerio de Ciencia, Investigaci\'on y Universidades (Spain) under the projects PID2019-110165GB-I00 (MCI/AEI/FEDER, UE) and PID2020-118758GB-I00, 
Ministerio de Ciencia e Innovaci\'on (Spain) MCIN/AEI/10.13039/501100011033/, Generalitat de Catalunya by the project 2017-SGR-929 (Catalonia), the EU STRONG-2020 project under the program  H2020-INFRAIA-2018-1 grant agreement no. 824093, and the German DFG through project no. 315477589 - TRR 211 (Strong-interaction matter under extreme conditions). This work was also partly supported by the Spanish programs Unidad de Excelencia Maria de Maeztu CEX2020-001058-M and CEX2019-000918-M, financed by MCIN/AEI/10.13039/501100011033.

\appendix

\section{Nonrelativistic limit of Euler and vorticity equations \label{app:NR}}

From the relativistic Euler equation~(\ref{eq:Euler})---neglecting any chiral imbalance---one can take the nonrelativistic limit by applying the expressions shown in Eq.~(\ref{eq:4velocity}). Expanding up  ${\cal O} (v^2/c^2)$ we obtain,
\be m (\pa_t + \bm{v} \cdot \bm{\nabla}) \bm{v} + \frac{1}{n}  \bm{\nabla} P = \bm{E} + \frac{\bm{v}}{c} \times \bm{B}  \ . \ee 
Then introducing the thermodynamic relation~(\ref{eq:dp}) without chiral imbalance, and the vector identity
\be (\bm{v} \cdot \bm{\nabla}) \bm{v} = \frac12 \bm{\nabla} v^2 - \bm{v} \times (\bm{\nabla} \times \bm{v}) \ , \ee
we get
\be m \pa_t \bm{v} + \bm{\nabla} \left( h_{\rm nr} + \frac{1}{2} m  v^2 \right) - m \bm{v} \times (\bm{\nabla} \times \bm{v}) = \bm{E} 
+ \frac{\bm{v}}{c} \times \bm{B} + T \bm{\nabla} \bar{s}
\ , \label{eq:Crocco}
\ee
which is the so-called Crocco equation~\cite{Gourgoulhon:2006bn,rezzolla} with an acceleration term coming from the Lorentz force.

\end{document}